\documentclass{emulateapj}
\usepackage{color}
\definecolor{orange}{rgb}{0.8,0.4,0}

\usepackage{url}
\bibstyle{aa}
\usepackage{color}

\shorttitle{Activity-Rotation of Mid-to-Late-type M Dwarfs}
\shortauthors{West et al.}
\submitted{Accepted for Publication in ApJ}

\begin{document}

\title{An Activity-Rotation Relationship and Kinematic Analysis of
  Nearby Mid-to-Late-type M Dwarfs}

\author{Andrew A. West\altaffilmark{1,2}, 
Kolby L. Weisenburger\altaffilmark{2,3},
Jonathan Irwin\altaffilmark{4}, Zachory
K. Berta-Thompson\altaffilmark{5},
David Charbonneau\altaffilmark{4},
Jason Dittmann\altaffilmark{4},
J. Sebastian Pineda\altaffilmark{6}}

\altaffiltext{1}{Corresponding author: aawest@bu.edu}
\altaffiltext{2}{Department of Astronomy, Boston University, 725
  Commonwealth Ave, Boston, MA 02215}
\altaffiltext{3}{Department of Astronomy, University of Washington,
  Box 351580, Seattle, WA, 98195}
\altaffiltext{4}{Harvard-Smithsonian Center for Astrophysics, 60 Garden St., Cambridge, MA 02138}
\altaffiltext{5}{MIT, Kavli Institute for Astrophysics and Space Research, 77 Massachusetts Ave., Bldg. 37, Cambridge, MA 02139}
\altaffiltext{6}{California Institute of Technology, Department of Astronomy, 1200 E. California Ave, Pasadena CA, 91125}

\begin{abstract}
  Using spectroscopic observations and photometric light curves of 238
  nearby M dwarfs from the MEarth exoplanet transit survey, we examine
  the relationships between magnetic activity (quantified by H$\alpha$
  emission), rotation period, and stellar age. Previous attempts to
  investigate the relationship between magnetic activity and rotation
  in these stars were hampered by the limited number of M dwarfs with
  measured rotation periods (and the fact that $v$~sin~$i$
  measurements probe only rapid rotation). However, the photometric
  data from MEarth allows us to probe a wide range of rotation periods
  for hundreds of M dwarf stars (from shorter than than one to longer
  than 100 days). Over all M spectral types that we probe, we find
  that the presence of magnetic activity is tied to rotation,
  including for late-type, fully convective M dwarfs. We also find
  evidence that the fraction of late-type M dwarfs that are active may
  be higher at longer rotation periods compared to their early-type
  counterparts, with several active, late-type, slowly rotating stars
  present in our sample. Additionally, we find that all M dwarfs with
  rotation periods shorter than 26 days (early-type; M1-M4) and 86
  days (late-type; M5-M8) are magnetically active. This potential
  mismatch suggests that the physical mechanisms that connect stellar
  rotation to chromospheric heating may be different in fully
  convective stars.  A kinematic analysis suggests that the
  magnetically active, rapidly rotating stars are consistent with a
  kinematically young population, while slow-rotators are less active
  or inactive and appear to belong to an older, dynamically heated
  stellar population.
  \end{abstract}


\section{Introduction}
Many late-type M dwarfs ($>$ M4) have strong magnetic fields that can
exceed 1 kG and heat their stellar chromospheres and coronae, creating
``magnetic activity'' that is observed from the radio to the X-ray
\citep[e.g.,][]{hawley96,west04,reiners08, berger08,reiners09,williams.2014.smomaudiaybn32j}.
Although magnetic activity has been observed in M dwarfs for decades,
the exact mechanism that gives rise to the chromospheric and coronal
heating is still not well-understood. The production of magnetic
fields and the subsequent activity may play a vital role in the
habitability of attending planets, which may be numerous given the
ubiquity of M dwarfs as planet hosts in the Galaxy
\citep[e.g.,][]{Charbonneau09,muirhead12,dressing13,dressing15}.  In addition,
given their fully convective interiors, late-type M dwarfs serve as
important laboratories for studying magnetic dynamo generation in
stellar (and potentially planetary) environments and are vital for
understanding the role that various stellar properties play in the
magnetic field and activity generation in low-mass stars.

In solar-type stars, magnetic field generation and subsequent heating
is closely tied to stellar rotation; the faster a star rotates, the
stronger its magnetic heating and surface activity.  Angular
momentum loss from magnetized winds slows rotation in solar-type
stars, and as a result, magnetic activity decreases with age. These effects
have been studied for decades, producing ample evidence for a strong
connection between age, stellar rotation, and magnetic activity in
solar-type stars \citep[e.g.,][]{skumanich72,barry88,
  soderblom91,barnes03,pizzolato03,mamajek08}.  All indications
suggest that this connection between age, rotation and activity
extends to early-type M dwarfs ($<$ M4), where rotation and activity
are strongly correlated \citep[e.g.,][]{pizzolato03, mohanty03,
  kiraga07}. The finite active lifetimes of early-type M dwarfs
observed in nearby clusters suggest that age continues to play an
important role in the rotation and magnetic activity evolution of
low-mass stars \citep{stauffer94, hawley99}.

At a spectral type of about M4, stars become fully convective \citep[0.35
M$_{\odot}$;][]{chabrier97,reid05}, a
property that may affect how magnetic field (and the resulting
heating) is generated.  Despite this change, magnetic activity
persists in late-type M dwarfs; the fraction of active M dwarfs peaks
around a spectral type of M7 before decreasing into the brown dwarf
regime \citep{hawley96, gizis00, west04}.  Several studies have
demonstrated that the large fraction of late-type M dwarfs observed to
be active is likely a result of their long activity lifetimes
\citep{silvestri05, west08}.

% this went from ~50,000 to the more precise 38,000, which assumes most of the W06 stars are included in the W08 sample
Using 37,845 M dwarfs from the Sloan Digital Sky Survey
\citep[SDSS;][]{york00}, \citet{west06,west08} found that stars
farther from the Galactic plane were less likely to be magnetically
active (as traced by H$\alpha$) than those near the Plane. They
interpreted the change in activity fraction as an effect of age: as
stars pass through the Galactic plane, they are dynamically heated by
gravitational interactions and obtain orbits that take them farther
from the Plane. Therefore, stars close to the Galactic plane are
statistically younger, while stars farther away are statistically
older.  \citet{west08} quantified this ``Galactic stratigraphy''
using a 1D dynamical model and derived the H$\alpha$ activity
lifetimes for M0-M7 dwarfs, finding that early-type M dwarfs (with
both radiative and convection zones) have active lifetimes of 1--2
Gyr, while late-type M dwarfs (with fully convective interiors) have
active lifetimes that exceed 7 Gyr. The level of activity (as
quantified by the ratio of the luminosity in H$\alpha$ to the
bolometric luminosity --- $L_{\rm{H}\alpha}$/$L_{\rm{bol}}$) also
appears to decrease as a function of stratigraphic age for all M
dwarfs (early and late-type), confirming that an age--activity
relation persists into the fully-convective regime \citep{west08}.

Tying activity (and age) to rotation for late-type M dwarfs has
been more challenging.  Recent simulations of dynamos find that
rotation may play a significant role in the magnetic field generation
of fully convective stars \citep{dobler06, browning08}.  From a
simple analytical model, \citet{reiners12} suggest that the
age-rotation relation extends to late-type M dwarfs and that the
angular momentum evolution of all stars is more a function of stellar
radius than it is interior structure.  Indeed, a few empirical studies
have uncovered evidence that activity and rotation might be linked in
late-type M dwarfs \citep{delfosse98, mohanty03, reiners08, browning10, reiners12}.  However, the majority of the observations have relied on
high-resolution spectroscopic data, which measure the rotational
velocity modified by the inclination of the star ($v$~sin~$i$).  While
$v$~sin~$i$ measurements provide important clues to the underlying
stellar rotation, they have a couple of limitations: 1) the best
spectrographs can produce $v$~sin~$i$ values only down to about 1
km\ s$^{-1}$, which for a 0.2 R$_{\odot}$ star corresponds to a
rotation period of 10 days, and are therefore completely insensitive to
slowly rotating stars; and 2) the inclination dependence adds
significant scatter to the derived rotation velocities.  Previous
results demonstrate these limitations by showing that almost all
late-type M dwarfs with a detected $v$~sin~$i$ have the same
``saturated'' level of magnetic activity; slowly rotating M dwarfs
with potentially less magnetic activity are not detectable.

An alternative method for studying stellar rotation is to use
photometrically derived rotation periods \citep[e.g.,][]{kiraga07}. Periodic signals result from brightness variations caused by
long-lived spots on the stellar surface rotating in and out of view.
While $v$~sin~$i$ observations require a single spectroscopic
observation, observing periodic, photometric variability requires
numerous, high-cadence observations that have been historically
prohibitive for large samples of late-type M dwarfs.  However, recent
programs to search for transiting planets around late-type M dwarfs
have produced large catalogs of time-domain photometry from which
can be gleaned several important stellar properties, including
rotation periods \citep{irwin09,law12}.

One of these transit programs, the MEarth
Project\footnote{\url{http://cfa.harvard.edu/MEarth/}}
\citep{nutzman08,irwinMearth,irwin11, berta12}, is employing two
arrays of robotic telescopes to photometrically monitor 4,000 nearby,
mid-to-late M dwarfs. MEarth data and associated observations have
provided new measurements of the fundamental properties of nearby
mid-to-late M dwarfs, including their distances \citep{dittmann14} and
their near-infrared spectra \citep[][]{newton14}.  Early in the
survey, \citet{irwinrot} determined photometric rotation periods
(ranging from 0.28 to 154 days) for 41 of the MEarth M dwarfs with
known parallaxes, and identified a relationship between these rotation
periods and kinematic age. As the mid-to-late M dwarfs observed by
MEarth are nearby and therefore relatively bright, it is possible to
obtain a larger sample of optical spectroscopy to probe magnetic activity, thus enabling a robust investigation of how the rotation-activity relation for fully convective M dwarfs extends to slowly rotating stars.

Determining ages for stars can be particularly challenging, especially
for low-mass stars \citep{soderblom10}.  In the past, age-dependent
relations for stars have been calibrated using stellar clusters.
However, the faintness of M dwarfs and the large distances to stellar
clusters with ages $>$ 1 Gyr preclude any detailed observations of
low-mass stars in the cluster environment.  One alternative method for
probing stellar age is to use the 3D stellar kinematics of a
population.  Historically, the dynamics of stars have been used to estimate
ages, as dynamical interactions cause the orbits of older stars to
become more elliptical and inclined to the Galactic plane \citep{wilson70, eggen89,leggett92}. Specifically, stars that
exhibit larger velocity dispersions have likely experienced more
dynamical encounters with molecular clouds and/or other stars and are
therefore older.  This is particularly true in the vertical or $W$
direction of stellar motion since most stars begin their lives in the
Galactic disk and slowly diffuse away via dynamical heating \citep{west06}.  Older stars also have slower azimuthal or $V$ mean
velocities due to asymmetric drift \citep[e.g.,][]{eggen89}.  While using
stellar kinematics to determine ages of individual stars is highly
problematic (since the dynamical processes that affect individual
stellar orbits are stochastic), 3D motions can reveal important age
information for bulk populations.  Dynamical analyses do require large
samples, which until recently have not been available for late-type M
dwarfs.

In this paper, we use an expanded sample of MEarth rotation periods for nearby M dwarfs, new optical spectroscopy of these targets, and their full
space motions to investigate the rotation-activity and
rotation-activity-age relation for late-type M dwarfs.  Section 2
describes the determination of rotation periods from MEarth data and the determination of magnetic activity from optical spectroscopic observations.  We describe the rotation, activity and kinematic
analysis in Section 3.  In Section 4 we demonstrate that there is a
close tie between rotation and activity and that the slowest rotators
come from a dynamically older population.  Lastly, Section 5
summarizes our conclusions and gives a brief discussion.

\section{Data}

\subsection{Rotation Periods from MEarth Photometry}\label{sec:rot}

To measure M dwarf rotation periods, we use photometric observations
from the MEarth survey for transiting exoplanets
\citep[see][]{nutzman08,berta12}. MEarth comprises two telescope
arrays, each consisting of eight robotic telescopes. All of the data
in this paper come from the MEarth-North array at the Fred Lawrence
Whipple Observatory (FLWO) at Mt. Hopkins, AZ, which has been
gathering data since 2008. The MEarth team recently commissioned a
duplicate array at Cerro Tololo Interamerican Observatory (CTIO),
called MEarth-South. Because the new array started regular
observations only in January 2014, data from MEarth-South are not
included in the following analyses. The MEarth M dwarf light curves
used in this paper (along with the full MEarth sample) are publicly available for download from the MEarth website\footnote{\url{http://cfa.harvard.edu/MEarth/Data.html}}.

The sample of M dwarfs observed by MEarth was designed to maximize sensitivity to the detection of small transiting planets ($2-4{\rm R_{\earth}}$), as detailed in \citet{nutzman08}. The MEarth sample is heavily biased toward mid-to-late M dwarfs within 33pc, with most stars having spectral types of M4-M6 but also including several stars of earlier and later spectral types. This paper makes use of the MEarth sample of stars and inherits the selection criteria of the planet-search survey. While the sample of stars observed was not designed explicitly to probe rotation across the entire M dwarf spectral class, it spans a sufficient range of spectral types to enable interesting comparisons between earlier (which in this paper refers primarily to M3-M4) and later (primarily M5-M6) spectral types.

The MEarth telescopes are 40 cm in diameter and equipped with back-illuminated CCDs. The telescopes observe automatically whenever conditions allow, with an adaptive scheduling algorithm selecting targets from a prioritized queue. The rotation periods in this paper are derived from ``planet cadence'' observations, in which a star is observed once every $10-30$ minutes for at least one observational season, resulting in light curves containing from hundreds up to a few thousand data points. For some stars, up to three observational seasons contributed to the rotation analysis. We took care to minimize changes to the telescopes to preserve long term stability, but those changes that were necessary are summarized in \citet{berta12} and documented in detail in the notes associated with the MEarth data releases.

The MEarth-North bandpass is fixed and nearly identical across the eight telescopes. Designed to maximize photon flux from red M dwarf targets, the bandpass is wide and has a shape set at short wavelengths by a longpass filter that cuts on at 715 nm and at long wavelengths by the quantum efficiency of the CCD, which decreases significantly by about 1000 nm. The time-variable optical depth through strong telluric water vapor features that fall within the MEarth bandpass can cause systematic photometric trends in MEarth M dwarf light curves, but the effect can be mitigated through an empirical correction \citep{berta12}. Photometric modulations due to rotating starspots are relatively muted in the MEarth bandpass compared to bluer wavelengths, but they can still have readily detectable peak-to-peak amplitudes of up to several percent \citep{irwinrot}. 

\begin{figure}[h]
\vspace*{-.3 cm}
\begin{center}
 \includegraphics[width=3.5in]{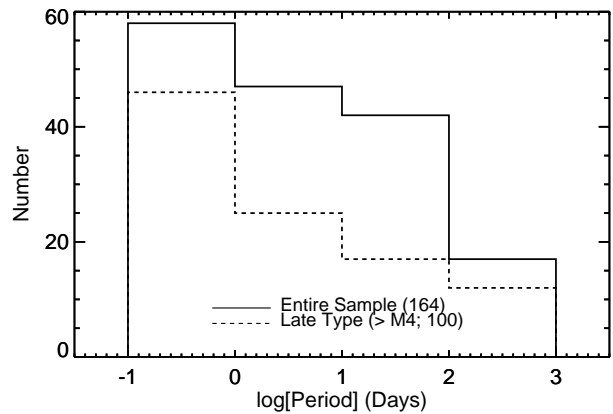} 
% \vspace*{-1.0 cm}
 \caption{ Distribution of rotation periods for all MEarth M dwarfs (solid)
   and late-type ($>$ M4) M dwarfs (dashed) with spectroscopic observations.}
   \label{fig:histo2}
\end{center}
\end{figure}

Rotation periods were extracted using a weighted least-squares
periodogram fitting algorithm that corrects for and marginalizes over
systematic photometric trends, including those caused by varying
telluric water absorption. This analysis is described in detail in
\citet{irwinrot}. Whereas \citet{irwinrot} included rotation periods only
for stars with published literature parallaxes, here we extend the
sample to all M dwarfs with rotational modulations detectable in the MEarth photometry that were available up to 2011, resulting in 164 targets.
For this paper's analysis, we applied a two-tier classification
system (``1'' and ``2'') for all of the measured rotation periods,
with ``1'' being the most robust detections.  Some of the reasons that
rotation periods were given a classification of ``2'' were if the stars had long
periods and less than two complete cycles were observed or if the stars had
larger uncertainties in the phase folding procedure. We were unable to
measure rotation periods for all of the MEarth stars. For stars with detected rotation period, the amplitude of the photometric modulations were typically $0.5-2\%$ peak-to-peak. \citet{irwinrot} made initial attempts to characterize the reliability and contamination of rotation period detection with MEarth and found the selection to be fairly clean for stars with $>$1\% variability, but messy (and strongly period-dependent) for stars showing variability of $0.5\%$ or less. Figure \ref{fig:histo2} shows the distribution of measured rotation periods for all of the M dwarfs with spectroscopic observations. The periods used in this paper can be found in Table \ref{table:data}.  A complete
catalog of rotation periods from the MEarth survey will be included in
a future publication (Newton et al. in prep).

\subsection{Magnetic Activity from FAST Spectroscopy}

To probe the magnetic activity of this sample, we obtained optical spectra for 238 MEarth M dwarfs. The sample of stars observed spectroscopically included those with measured rotation periods, as well as additional M dwarfs for which no rotational modulations were detected.  Observed using the 600 lines/mm grating on the FAST spectrograph on the 1.5m Tillinghast Telescope at FLWO, the spectra have a resolution of $R=3000$.  The spectra cover 5550-7550 \AA\ and include several features that are important for M dwarf analyses, including (but not limited to) the CaH and TiO molecular bands and the H$\alpha$ atomic line. 

The FAST observations were acquired over 30 nights from December 2010 to
July 2012, with HeNeAr lamp exposures taken at every telescope
position and spectrophotometric standards observed every
night. Exposures lasted typically about five minutes, yielding
signal-to-noise ratios (SNR) of about 50 per resolution element. With a scripted IRAF reduction, we bias-subtracted and flat-fielded all spectra using calibration exposures taken every time the instrument grating changed, and applied a wavelength calibration determined for each spectrum from its matching HeNeAr exposure. We extracted one-dimensional spectra for each target, weighting by the profile in the cross-dispersion direction and using linear interpolation to subtract the sky. We determined a rough flux calibration for each spectrum using the nightly spectrophotometric standards.

\subsection{Distances from MEarth Astrometry and Other Sources}
To characterize the basic properties of the target M dwarfs, we used
distance measurements from several sources. \citet{dittmann14} used MEarth imaging
to measure geometric parallaxes for 1,507 M dwarfs in the MEarth
sample. These include 150 out of the 164 stars in the rotation sample
and 213 out of the 238 M dwarfs in the spectroscopic sample. For the remaining stars without MEarth parallaxes, we adopted distances published in \citet{lepine05}, which come either from literature trigonometric parallaxes or from photometric parallax estimates.  Proper motions for all of the stars were measured as part of the LSPM-North catalog \citep{lspm}.

\section{Analysis}

\begin{figure}[h]
\vspace*{-.3 cm}
\begin{center}
 \includegraphics[width=3.5in]{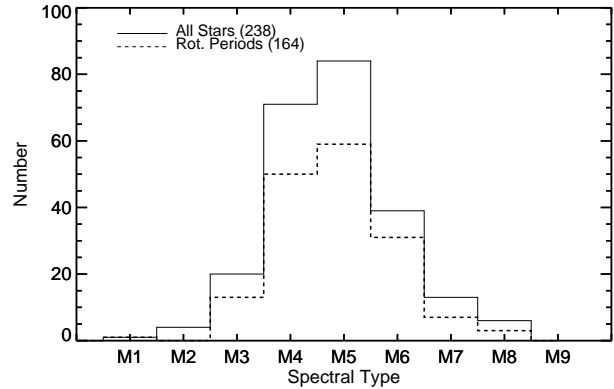} 
 %\vspace*{-1.0 cm}
 \caption{Distribution of spectral types for MEarth stars with FAST spectroscopic
   observations, including stars with measured rotation periods
   (dashed).}
   \label{fig:histo1}
\end{center}
\end{figure}

The FAST spectra were processed with the Hammer spectral-typing
facility \citep{covey07}.  The Hammer uses
measurements of atomic and molecular features to estimate an initial
spectral type. We then visually inspected all of the spectra with the
Hammer (v.\ 1\_2\_5) ``eye check'' function, and manually assigned
spectral types to each star -- we identified and classified 238 M
dwarfs.  (see Figure \ref{fig:histo1}).  Of the 238 M dwarfs for which
we obtained FAST spectroscopy, 164 have measured rotation periods.

We measured radial velocities (RVs) by cross-correlating each FAST
M dwarf spectrum (between 5600-7100 \AA) with the respective \citet{bochtem} M dwarf template \citep{mohanty03,westbasri09}.  We
determined the typical uncertainty in our radial velocity by comparing
our calculated RVs for 11 stars with previously determined RVs from
high-resolution spectra
\citep{delfosse98,montes01,nidev02,mohanty03,browning10,shkolnik10}. The
RMS difference between the RV measurements was 5.1 km\ s$^{-1}$, which
we took as the uncertainty in our ability to measure RVs from the FAST
spectra.   Each spectrum was corrected to air wavelengths in a
zero-velocity heliocentric rest frame for the measurements of spectral
features (see below).

\subsection{Magnetic Activity}\label{sec:act}

Using the radial velocity-corrected spectra, we determined the presence and
magnitude of magnetic activity by measuring H$\alpha$ equivalent
widths (EWs) as defined by \citet{west11}. We used 6500-6550 \AA\ and
6575-6625 \AA\ as our two continua regions, with the central H$\alpha$
wavelength at 6562.8 \AA.  In previous studies, stars with H$\alpha$
EWs greater than 0.75 \AA\ were considered magnetically
active \citep{west11}. However, given the relatively small size of our
FAST sample, we were able to visually inspect each spectrum and
classified all stars with detectable H$\alpha$ emission as
``magnetically active.''  All of the active stars in our sample have
H$\alpha$ EWs $>$ than the 0.75 \AA\ criterion from \cite{west11}, and
would have been selected as ``active'' using an automatic EW threshold.
The high SNR of the
FAST spectra allowed us to be sensitive to H$\alpha$ EWs far below the
values measured for the active stars in our analysis (particularly for the M3-M6 dwarfs), indicating that
we could have detected activity in many of the ``inactive'' stars had
it been present.  

M dwarfs that exhibit weak magnetic activity can show H$\alpha$ in absorption rather than emission \citep{stauffer86, walkowicz09}. As such, using H$\alpha$ emission as an activity indicator will not include M dwarfs with the weakest levels of magnetic activity. Throughout this paper, we define active stars as those that exhibit H$\alpha$ in emission and are therefore strongly magnetically active. We examined all of the inactive early-type M dwarf spectra to look for H$\alpha$ absorption, and found no evidence for strong absorption. In 9 of 47 of these cases, there were hints of weak H$\alpha$ absorption. However, a strong nearby TiO absorption feature precluded firm detections, given the low resolution of the FAST spectra. Activity fractions reported in this paper may be slightly higher in reality, because we may have excluded weakly active M dwarfs that do not show H$\alpha$ in emission.

\begin{figure}[h]
\vspace*{-0.3 cm}
\begin{center}
 \includegraphics[width=3.5in]{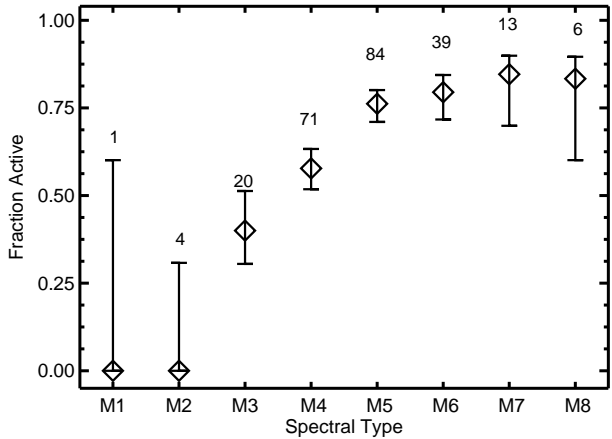} 
 %\vspace*{-1.0 cm}
 \caption{The fraction of magnetically active M dwarfs as a function of
   spectral type.  Errors were computed from binomial statistics and
   the numbers above each symbol indicate the total number of stars
   (active + inactive)  in each bin.  Similar to previous results, we
   see a dramatic increase in the fraction of active M dwarfs with
   increasing spectral type.}
   \label{fig:fracact}
\end{center}
\end{figure}

We identified 160 active M dwarfs, which
comprises 67\% of the M dwarfs in the spectroscopic
sample. Figure \ref{fig:fracact} shows the fraction of magnetically
active M dwarfs in our sample as a function of spectral type.  As seen in previous
studies \citep{hawley96, gizis00, west04}, we
see an increase in activity with increasing spectral type, which is
likely due to the longer active lifetimes of late-type M dwarfs
\citep{west08}.

There were previously determined H$\alpha$ EW values for 23 of the
stars in our sample \citep{reid95}. We compared the FAST EWs to those
from \citet{reid95} and found that 12 of the measurements are
within 13\% of those previously reported with the median difference
being 14\% for all stars. Six of the stars have EWs that are
different by more than 25\% and only two show variations larger than 50\%.
Variation on this level can be explained by the typical variation of
H$\alpha$ emission in M dwarfs \citep[e.g.,][]{bell12}. With the activity criterion used in this paper, none of the 23 stars switch from being active to being inactive (or vice versa) between the two epochs.

\begin{figure}[h]
\vspace*{-0.3 cm}
\begin{center}
 \includegraphics[width=3.5in]{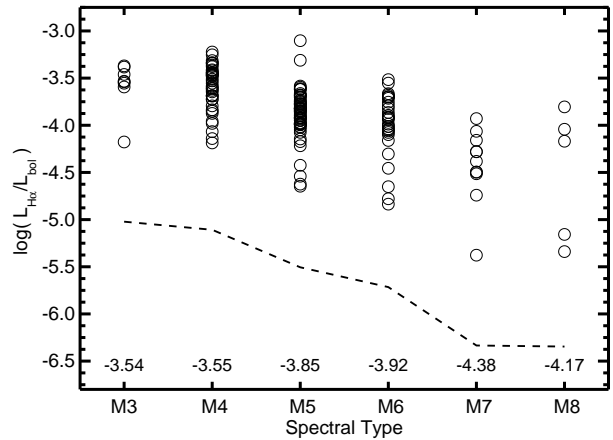} 
 %\vspace*{-1.0 cm}
 \caption{$L_{\rm{H\alpha}}$/$L_{\rm{bol}}$ as a function of spectral
   type for active stars in the sample. Spectral types earlier than M3 are not shown, as our sample contains no such stars that are active. Median values of log($L_{\rm{H\alpha}}$/$L_{\rm{bol}}$) are listed below.   The dashed line indicates the typical
  $L_{\rm{H\alpha}}$/$L_{\rm{bol}}$ values to which we are
  sensitive as a function of spectral type. There is a large spread in the
   $L_{\rm{H\alpha}}$/$L_{\rm{bol}}$ values at each spectral type, as
   well as a significant decrease in the median
  $L_{\rm{H\alpha}}$/$L_{\rm{bol}}$ values with increasing spectral
  type.  }
   \label{fig:fracact2}
\end{center}
\end{figure}

For the active stars in the sample (those with H$\alpha$ emission), we
calculated the magnetic activity strength,
$L_{\rm{H\alpha}}$/$L_{\rm{bol}}$ using the $\chi$ factor from
\citet{walkowicz04}. We multiplied the spectral
type-dependent conversion factor $\chi$ by the H$\alpha$ EW to
get the ratio $L_{\rm{H\alpha}}$/$L_{\rm{bol}}$.  Figure
\ref{fig:fracact2} shows the $L_{\rm{H\alpha}}$/$L_{\rm{bol}}$ values
as a function of spectral type.  The dashed line in Figure
\ref{fig:fracact2} indicated the typical level of H$\alpha$ emission
to which we are sensitive as a function of spectral type.  This
detection threshold was estimated using the typical noise level near H$\alpha$ in the FAST
spectra and assuming a detectable emission line would have a peak
that is at least three times larger than the noise. As shown in previous studies
\citep[e.g.,][]{burgasser02, west04}, we see a decrease in
$L_{\rm{H\alpha}}$/$L_{\rm{bol}}$ with increasing spectral type (well
above detection threshold for M3-M6 dwarfs; dashed line) as
well as a large spread of $L_{\rm{H\alpha}}$/$L_{\rm{bol}}$ at each
spectral type.  The latter is a combination of the intrinsic spread of
the active M dwarf population and age-dependent effects on the
activity that are washed out when stars of different ages are binned
together.

As part of our analysis, we also measured the TiO, CaH, and CaOH molecular
bandheads as defined by \citet{reid95}, and the NaI atomic absorption line. All
of the values can be found in the online version of Table \ref{table:data}.

\subsection{Kinematics}

We combined the stellar positions, proper motions, distances and
radial velocities to compute the Galactic $U$, $V$, $W$ velocities for 212
of the stars in our sample.\footnote{We used a right-handed coordinate
  system with $U$ pointed towards the Galactic center.}  We corrected
for the solar motion with respect to the Local Standard of Rest \citep[11,
12, 7 km\ s$^{-1}$;][]{schonrich10} and propagated our measurement
errors through to formal uncertainties in $U$, $V$, and $W$ using a
modified version of the IDL routine {\tt GAL\_UVW.PRO}.

There were 159 stars in our sample that had both good rotation
measurements (flags ``1'' or ``2'' from Section 2) and 3D space motions.  We divided these stars into four
different bins based on the rotation velocities ($<$1 days, 1--10
days, 10--100 days, and $>$100 days; with 58, 47, 42 and 17 stars in
each bin respectively). 

The velocity distributions of specific stellar populations can be
modeled by Gaussian distributions in all three kinematic components.
Mean space velocities and their respective spreads and uncertainties
are often calculated using fits to cumulative probability
distributions or simple Gaussian fits to velocity histograms
\citep{reid95, bochUVW, reiners09}.  However, these methods are
computationally intensive and/or heavily biased by the binning
strategy.  The challenges associated with previous methods are
amplified when the data are sparse, which is particularly important
for our small number of slow rotators.

Alternatively, under the assumption that the underlying velocity
distribution is Gaussian, probabilities for a given model mean and
standard deviation can be computed analytically and compared to the
data using a Bayesian approach.  For a given subset of velocities, the
joint probability distribution for the best fit mean ($\mu$) and
standard deviation ($\sigma$) of the population is given by:

\begin{equation}
p(\{ \mu,\sigma \}; \{ {\rm data} \}) \propto p(\{ {\rm data} \};\{ \mu,\sigma \}) p(\{ \mu,\sigma \}),
\end{equation}

\noindent where the first term on the right represents the likelihood of the
data and the second term on the right is the prior distribution of
parameters, which we took to be uninformative.\footnote{We used the Jeffreys prior, 1/$\sigma$, also known as the logarithmic prior, for the standard deviation and a flat prior for the mean.}

The likelihood is the product of the probabilities for all of the
data points in a subset, where the probability for each datum is given
by the Gaussian distribution:

%\begin{equation}
\begin{eqnarray}
p(\{ m,s \}_i;\{ \mu,\sigma \}
)=&\frac{1}{\sqrt{2\pi(\sigma^2+s_i^2)}}\exp{\left[
    \frac{-(m_i-\mu)^2}{2(\sigma^2+s^2_i)}\right]},
\end{eqnarray}
%\end{equation}
 
\noindent where $m$ and $s$ are the measured velocity and uncertainty of a
single star, $p$ is the probability that the datum comes from a
Gaussian distribution described by $\mu$ and $\sigma$, and the index
$i$ runs over all stars within a bin if it contains at least three stars. 

We constructed a grid of $\mu$ and $\sigma$ values and explored the
joint probability distribution of Equation 1 for the four
kinematic sub-samples at each of the grid points.  From the resulting peaks
of the joint probability density distributions, we
computed $\mu$ and $\sigma$ values for each
rotation/activity sub-sample.  We also calculated realistic uncertainties from
the spread of the marginalized distributions for each parameter. The results of this
analysis can be found in Table \ref{table:kinematics} and are
discussed below in more detail. For the fastest rotators, the data did
not provide a good constraint on the $U$ velocity dispersion; that
value was excluded from Table \ref{table:kinematics}. 

\section{Results}

\subsection{Rotation vs. Activity}

\begin{figure}[h]
\vspace*{-0.3 cm}
\begin{center}
 \includegraphics[width=3.5in]{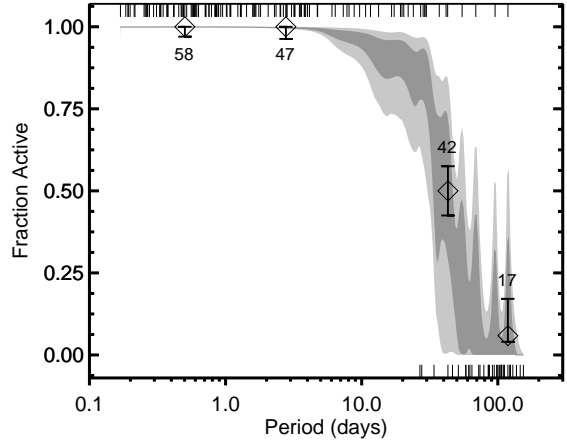} 
%\vspace*{-1.0 cm}
 \caption{The fraction of magnetically active stars for all M dwarfs in our
   sample as a function of rotation period.  Diamonds represent
   logarithmically-spaced bins with error bars
   calculated from binomial statistics. The numbers above/below each
   symbol represent the number of stars in each bin. The gray shaded
   regions show the nonparametric probability density function of the
   activity fraction with
   confidence intervals (dark gray - 68\%; light gray - 95\%)
   determined from 5000 bootstrap samples of the data.  Rug plots on
   the top and bottom indicate the active and inactive (respectively) stars used to
   compute the activity fraction. All of
   the fast rotators are active, whereas a small but non-zero fraction of the
   slow rotators show magnetic activity.}
   \label{fig:peract}
\end{center}
\end{figure}

From the sample of 164 M dwarfs with measured rotation periods, 127
were classified as magnetically active.  We investigated how the
activity fraction varies as a function of rotation period by dividing
the sample into four logarithmically spaced rotation bins ($<$1 days, 1--10
days, 10--100 days, and $>$100 days).  We also used the Adaptive
Kernel Density Estimation routine {\tt akj} within the
{\tt quantreg} package in R to construct probability density functions (PDF)
for the active and inactive stars as a function of rotation period \citep{r,quantreg}. We then combined
the active and inactive PDFs to compute a PDF for the
activity fraction as a function of rotation period. We determined
confidence intervals (68\% and 95\%) by taking 5000 bootstrap samples
of the data. Figure
\ref{fig:peract} shows the activity fraction as a function of
rotation period for the logarithmic bins (diamonds) and the
nonparametric PDF (gray shading).  Figure
\ref{fig:peract} indicates that all M dwarfs with periods shorter than ten days show detectable H$\alpha$ emission, with a decrease in activity
fraction as a function of increasing rotation period.  While the
slowest rotating M dwarfs have lower activity fractions, there appears
to be a small population of active, slowly rotating stars.

\begin{figure*}
\vspace*{-0.3 cm}
\begin{center}
 \includegraphics[width=6in]{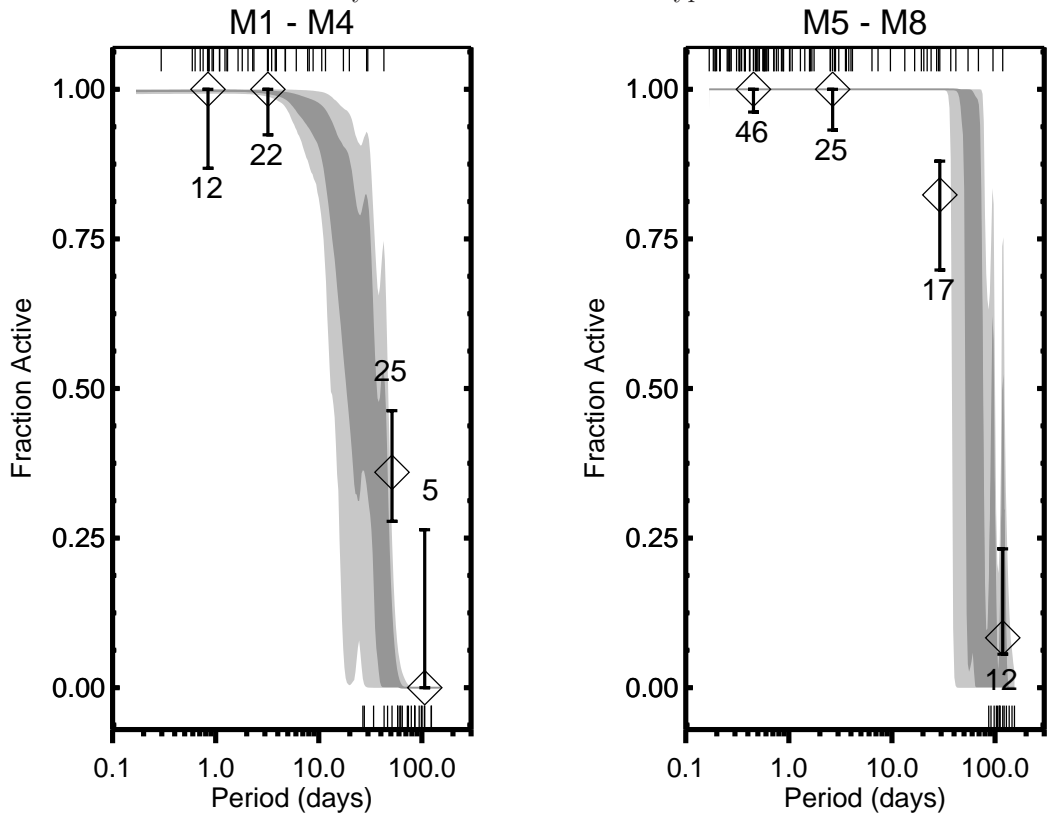} 
%\vspace*{-1.0 cm}
 \caption{The fraction of active early-type (left) and late-type
   (right) M dwarfs in our sample as a function of rotation period.
   Diamonds represent logarithmically-spaced bins with error bars
   calculated from binomial statistics. The numbers above/below each
   symbol represent the number of stars in each bin. The gray shaded
   regions show the nonparametric probability density function of the
   activity fraction with confidence intervals (dark gray - 68\%;
   light gray - 95\%) determined from 2000 bootstrap samples of the
   data.  Rug plots on the top and bottom indicate the active and
   inactive (respectively) stars used to compute the activity
   fraction. At similar rotation periods, a much larger fraction of
   the late-type M dwarfs are active, evidencing that the
   activity-rotation relation in M dwarfs may be mass dependent.}
   \label{fig:peract2}
\end{center}
\end{figure*}

We further explored the rotation dependence on activity fraction by
dividing the sample into early (M1--M4) and late-type (M5--M8)
samples. Figure \ref{fig:peract2} was made using the same
procedures as in Figure \ref{fig:peract} except that we used 2000 bootstrap
samples to determine the PDF confidence intervals. Figure
\ref{fig:peract2} demonstrates a clear difference in the populations
with the early-type M dwarfs (left) showing a strong decrease in
activity fraction with increasing rotation period, with the slowest
rotators being completely inactive.  The activity fraction of the
late-type population (right) stays very high for stars with
rotation periods shorter than 100 days, after which it decreases.
A small number of the slowest rotating, late-type M dwarfs
still show magnetic activity.  While differences in the activity
fractions between early and late-type M dwarfs have been seen for
decades \citep[e.g.,][]{gizis00,west04} and can be explained by
differences in active lifetimes \citep{west08}, Figure
\ref{fig:peract2} suggests that the relationship between activity and
rotation may be different in late-type versus early-type M dwarfs.
Specifically, it appears that late-type M dwarfs may stay active at
slower rotation speeds than their early-type counterparts.

\begin{figure*}[h]
\vspace*{-0.3 cm}
\begin{center}
 \includegraphics[width=6in]{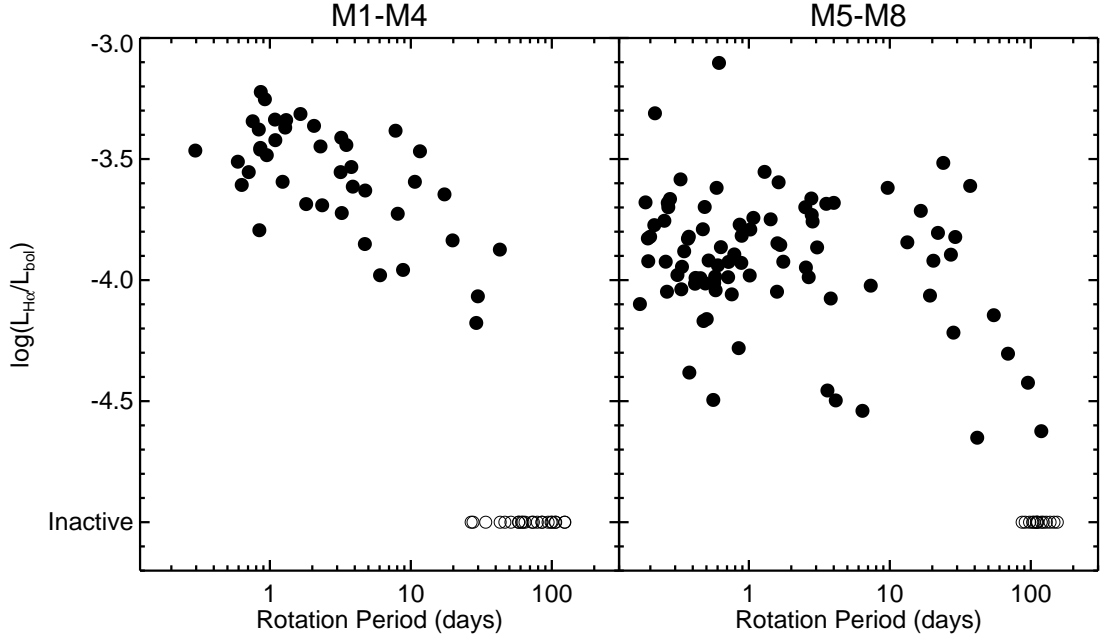} 
%\vspace*{-1.0 cm}
 \caption{log($L_{\rm{H\alpha}}$/$L_{\rm{bol}}$) as a function of rotation
   period for early-type (left) and late-type (right) M dwarfs (filled
   circles).  Inactive stars with measured rotation periods are
   included as open circles.  In the early-type M
   dwarfs there is a clear decease in the strength of magnetic activity with
   increasing rotation period.  The late-type
   M dwarfs are much more scattered, but appear to maintain a similar
   level of activity except for the very slowest rotators.  The
   inactive stars suggest that there may exist rotation periods (26 days; early-type and 86 days; late-type), faster
   than which all stars are
   magnetically active. }
   \label{fig:actrot_nobin}
\end{center}
\end{figure*}

\begin{figure*}[h]
\vspace*{-0.2 cm}
\begin{center}
 \includegraphics[width=6in]{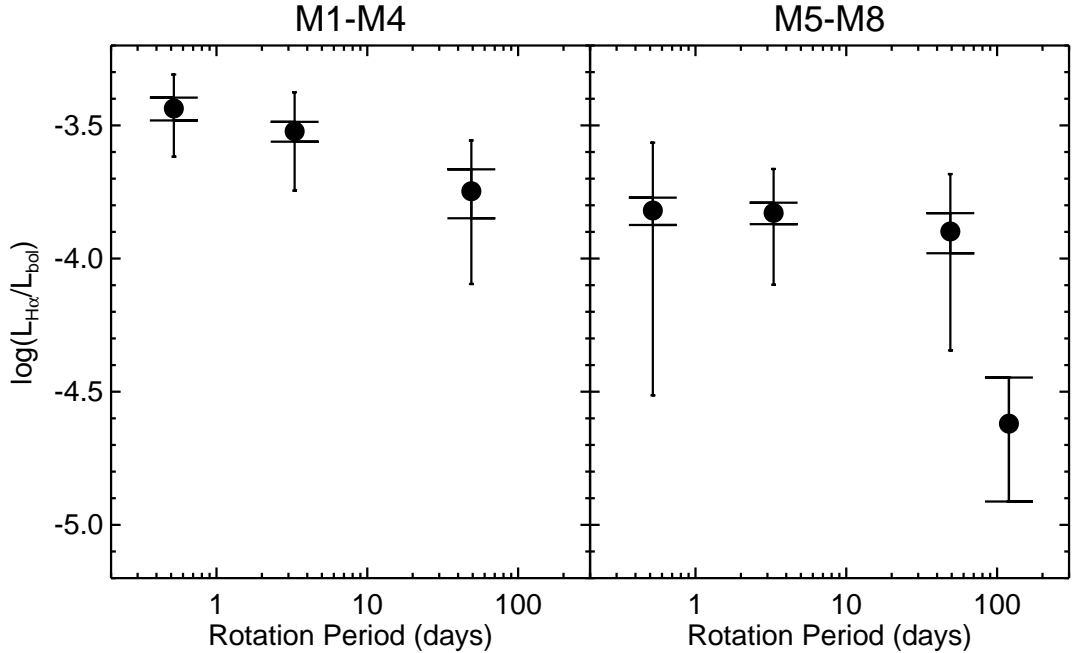} 
 %\vspace*{-1.0 cm}
 \caption{log($L_{\rm{H\alpha}}$/$L_{\rm{bol}}$) as a function of rotation
   period for early-type (left) and late-type (right) M dwarfs.  Each
   data point corresponds to the mean value (of activity and rotation
   period) for the stars in the bin.  The vertical error bars
   represent the spread of the data (thin bars) and the uncertainty
   in the mean values (wide bars). Only three bins are shown for the
   M1-M4 dwarfs, as no active stars of these types were found with
   periods longer than 100 days. There is only one star in the slowest
   rotating late-type bin and the error bars on this data point represent the
   uncertainty in the $L_{\rm{H\alpha}}$/$L_{\rm{bol}}$ measurement.
   In the early-type M
   dwarfs there is evidence of a decease in the strength of magnetic activity with
   increasing rotation periods.  This is in contrast to the late-type
   M dwarfs, which
   maintain a similar level of activity except for the very slowest rotators.}
   \label{fig:actrot}
\end{center}
\end{figure*}

We also investigated how the strength of magnetic activity (quantified
by $L_{\rm{H\alpha}}$/$L_{\rm{bol}}$) varies as a function of rotation
period. Figure \ref{fig:actrot_nobin} shows
log($L_{\rm{H\alpha}}$/$L_{\rm{bol}}$) (filled circles) as a function of rotation
period for early-type (left; M1--M4) and late-type (right; M5-M8) M
dwarfs. The inactive stars with measured rotation periods are plotted
as open circles. The early-type M dwarfs show a decrease in activity strength with increased rotation
period. In contrast, the late-type M dwarfs are consistent with having the same
level of activity, except perhaps at the longest rotation
periods.  To quantify this, we performed a linear least squares fit to the active stars in both panels of Figure \ref{fig:actrot_nobin}. In the early-type dwarfs, log($L_{\rm{H\alpha}}$/$L_{\rm{bol}}$) shows a statistically significant decrease as function of rotation period, and has the form,

\begin{eqnarray}
\log(L_{\rm{H\alpha}}/L_{\rm{bol}}) &= & -3.44 \pm 0.02 \nonumber \\ 
                                      & & - 0.19 \pm 0.04\times \log(P_{\rm rot}),
\end{eqnarray}

\noindent with $P_{rot}$ as the rotation period in days. For the late-type M dwarfs, log($L_{\rm{H\alpha}}$/$L_{\rm{bol}}$) shows no significant trend with rotation period (slope = $-0.016 \pm 0.050$). The slope of the active late-type dwarfs is not consistent with the decreasing trend seen in the active early-type M dwarfs. In both cases, the scatter about the best-fit line is dominated by the intrinsic scatter rather than measurement uncertainty.
Binning the data in logarithmic bins of rotation period confirms these
bulk trends (Figure \ref{fig:actrot}).  Both Figures
\ref{fig:actrot_nobin} and \ref{fig:actrot} demonstrate that there is
a connection between rotation and activity in late-type M dwarfs; all
stars rotating with periods shorter than $\sim$90 days are active.  There is also a hint that there may be a
correlation between rotation and activity in the late-type M dwarfs
with rotation periods longer than 90 days, but there are not enough stars to make any strong
conclusions (e.g., the slowest rotation bin of Figure \ref{fig:actrot}
contains only one star).  Additional late-type, slowly rotating, active M
dwarfs are required to confirm if magnetic activity is weaker in slowly rotating, fully
convective M dwarfs.

Figure \ref{fig:actrot_nobin} also suggests that while fast rotation
is correlated with magnetic activity in all M dwarfs, there are clear
differences between the early and late-type populations.  Similar to what was seen in Figure
\ref{fig:peract2}, the late-type M dwarfs maintain a high (and
comparable) level of magnetic
activity until their rotation periods exceed $\sim$90 days, beyond which
the level of activity appears to decrease.  In the early-type M
dwarfs,  there is a much more clear decrease in
activity at all
rotation periods and an absence of active stars with periods $>$ 30
days.  The inactive stars in both panels corroborate that rotation plays an
important role in the generation of magnetic activity in M dwarfs and
hint at the possibility of a rotation threshold faster than which
magnetic activity is present.  All M dwarfs rotating
faster than 26 days and 86 days in the early-type and late-type
populations respectively are magnetically active.

\begin{figure}[h]
\vspace*{-0.3 cm}
\begin{center}
 \includegraphics[width=3.5in]{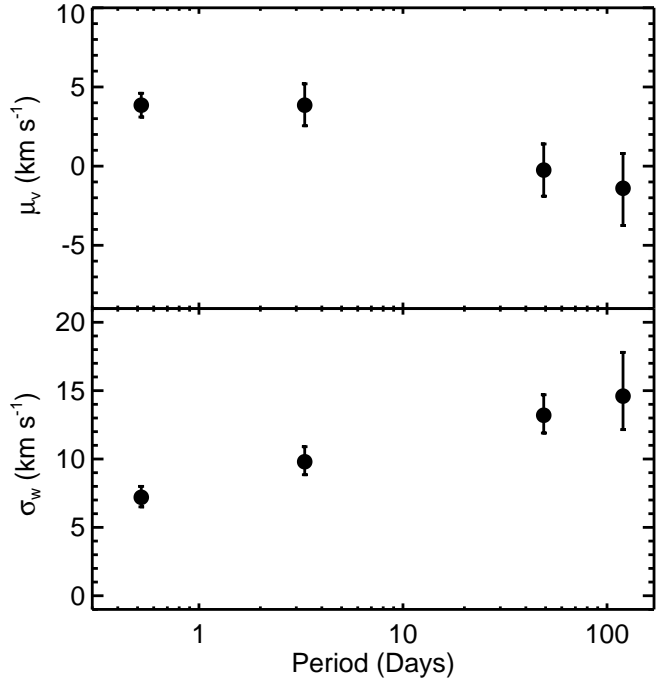} 
 %\vspace*{-1.0 cm}
 \caption{The mean $V$ velocities (top), and $W$ velocity dispersions
   (bottom) as a function of rotation period.  Error bars represent
   the 1$\sigma$ uncertainties in each bin. Both panels suggest that
   slowly rotating M dwarfs are drawn from an older kinematic population.}
   \label{fig:velmean}
\end{center}
\end{figure}

Figures \ref{fig:actrot_nobin} and \ref{fig:actrot} also demonstrate an alternative method for
studying the range of magnetic activity strengths in M dwarfs.  Our
results reduce the contribution of different rotation periods to the
spread of activity strength and allowed us to examine the intrinsic
spread of the population (at a given rotation rate), particularly for
early-type M dwarfs.  While the mean
values have relatively small uncertainties (wide error bars) in most
bins, the spread (thin error bars) at each rotation bin remains large
(typically about 0.3--0.5 dex).  We see that mixing stars of various
rotation rates (and perhaps different ages) \emph{does} increase the
spread of the population (particularly for early-type M
dwarfs). Additionally, some
of the spread seen in individual bins in Figure
\ref{fig:actrot} indicate that the M dwarf population has a
significant intrinsic variety of activity strengths.  As has been seen
in previous studies \citep[e.g.,][]{berger08,lee12,bell12}, much of
the observed spread may be due to the H$\alpha$ variability of
individual M dwarfs.

\subsection{Kinematics, Rotation and Activity}

From the results of the kinematic analysis described in Section 3.2,
we examined the bulk motions of the sample as a function of rotation
period.  Figure \ref{fig:velmean} shows the mean $V$ velocity (top)
and the $W$ velocity dispersion (bottom) of the entire sample
(relative to the Local Standard of Rest) as a function of rotation
period. The $V$ component of the velocity (in the direction of
Galactic rotation) shows clear signs of increasing asymmetric drift
with increasing rotation period.  Namely, the mean $V$ velocity
decreases for stars with slow rotation rates.  This decrease is a
strong indicator that the population of slow rotators is older;
dynamical heating over time causes elliptical orbits, which lead to a
bulk decrease in $V$ motion. The mean $U$ and $W$ motions are
insensitive to asymmetric drift and provide little information about
the ages of small samples of stars.  All of the population $U$, $V$, and $W$
values (means and dispersions) are included in Table
\ref{table:kinematics} for completeness.

The velocity dispersion of the $W$ component (bottom panel of Figure
\ref{fig:velmean}; in the vertical direction, out of the plane of the
Galaxy) increases with increasing stellar rotation period.  Increases
in the vertical velocity dispersion are typically associated with
multiple dynamical encounters and older ages.  As with the asymmetric
drift apparent in the mean velocities, the vertical velocity
dispersion measurements suggest that the slower rotators are
dynamically older. The dispersions of the $U$ and $V$ components can
be affected by other processes and are less directly linked to
dynamical heating than the $W$ velocity dispersions, but are also included
in Table \ref{table:kinematics}.

% \begin{figure}[h]
% \vspace*{-0.3 cm}
% \begin{center}
%  %\includegraphics[width=4.2in]{../veldisp_all.eps} 
% % \vspace*{-1.0 cm}
%  \caption{The $U$ (top), $V$ (middle), and $W$ (bottom) velocity dispersions as a
% function of rotation period for active stars (gray triangles) and the entire sample
% (black circles). Error bars represent the  1$\sigma$ uncertainties in each bin. }
%    \label{fig:veldisp}
% \end{center}
% \end{figure}

We also completed the same kinematic analysis for
separated late-type and early-type populations and found no significant 
differences in their kinematics.

\section{Conclusions and Discussion}

The spectroscopic follow-up of 238 M dwarfs, combined with the rotation
rates from the MEarth transit survey has
provided an unprecedented sample for examining the rotation,
magnetic activity and kinematics of mid-to-late M dwarfs.  We conducted an
analysis of the magnetic activity properties and investigated the bulk
3D kinematics of the population  as a function of rotation.  In the spectroscopic sample,
164 stars have measured rotation periods, 160 are magnetically
active, and 127 of the M dwarfs are both active and have measured
rotation periods.  We provide all of the sample and derived quantities
in the electronic version of this manuscript.  

The results of our analysis are summarized as follows:

\begin{enumerate}

\item For early-type M dwarfs, the activity fractions decrease with
increasing rotation period.  For the late-type M dwarfs, the higher
activity fractions extend to slower rotation rates before rapidly declining.

\item All of the rapidly rotating stars are magnetically active, with
few (but some) of the slowest rotators showing signs of activity. 

\item For early-type (M1-M4) M dwarfs, all stars rotating faster than
  26 days are magnetically active, as are all late-type M
  dwarfs (M5-M8) with rotation periods shorter than 86 days. It is
  worth noting that for a 0.2 M$_{\odot}$ star, 26 and 86 days
  correspond to $v$~sin~$i$ values of 0.4 and 0.1 km\ s$^{-1}$ respectively, both
  of which are below the detection limits of most spectrographs.

\item At the longest rotation periods, we see no active early-type M dwarfs,
but there are a few slowly rotating, active, late-type M dwarfs.

\item The strength of magnetic activity appears to decline with increased rotation
period in early-type M dwarfs.  The fully-convective, late-type M
dwarfs remain at a similar activity level as rotation slows, perhaps until the slowest
rotation periods ($>$ 90 days).

\item From our kinematic analysis, it appears that M dwarfs with longer
rotation periods are drawn from a kinematically older population.
Specifically, we see that slowly rotating stars exhibit signs of
asymmetric drift and have larger velocity dispersions about the mid-plane, both signs of
dynamically heated, older populations.

\item There are a few slowly rotating M dwarfs that show signs of
  magnetic activity, and which may provide important laboratories for
  additional investigations about the role rotation plays in the magnetic
  field generation in low-mass stars.

\end{enumerate}

In general, we find a strong relationship between rotation and
activity for early-type M dwarfs.  The relationship is much less
pronounced, but appears to
exist for the late-type M dwarfs that are
fully convective.  For all M dwarfs, rotation appears to
affect the presence of activity; fast rotators are more likely to show strong activity, while slow rotators
are less likely to be active.  However, in the late-type M dwarfs,
magnetic activity appears to be present in stars with slower rotation
than their early-type counterparts.  Larger samples that include
measurements of magnetic activity and rotation are required to further
investigate this discrepancy (e.g. Newton et al. in prep).

Our kinematic analysis suggests that rotation is linked to age
in all M dwarfs.  We conclude that younger M dwarfs
(both early and late-type) are rotating
faster and more likely to be active.  Future analysis of wide binaries
(particularly those with a white dwarf companion from which we can
determine age) and/or stellar
clusters will provide better information about the detailed role that
stellar age plays in the observed rotation and magnetic activity of
low-mass stars.

While our kinematic analysis
shows clear signs of asymmetric drift and dynamical heating, the
magnitudes of the effects are far less than what is seen in large samples of M
dwarfs that probe farther into the Milky Way disk
\citep[e.g.,][]{bochUVW}.  This suggests that the slow rotators in our
sample are older than the fast rotators, but that they are still
younger than typical old stellar populations in the Galactic (thin) disk.  This is
not surprising given that the MEarth targets are selected to be nearby, where
it is much more likely to sample a younger population \citep{west08}.

Some of the active stars in our spectroscopic sample do not have
measured rotation periods (21\%) and were therefore not included in
some of our analyses. 74 stars in the sample do not have
rotation periods (55\% of which are inactive).  Because our ability to
measure a rotation rate of an M dwarf depends on the presence of starspots, it is not surprising that some stars, namely the inactive ones,
may not show rotational modulation; indeed, the majority of the stars
with rotation periods are active (77\%). However, the lack of measured
rotation in the active stars could be due to a number of effects,
including but not limited to, homogeneous spot geometry, short-lived
spots, insensitivity to low-amplitude variations (few small spots), or
an over abundance of spots.  The latter effect would cause a decrease
in the rotational modulation for the most active (heavily spotted)
stars. We ran the non-rotating stars through our kinematic analysis
(53 have 3D kinematics) and found that as a population they are
consistent with M dwarfs in the 10-100 day rotation bins.  We conclude that kinematically, the stars without rotation measurements
could be on the older end of the spectrum for our sample.  The active fractions
of early and late-type M dwarfs without rotations measurements are
consistent with the entire sample in the 10-100 day rotation
bins.  Therefore, the exclusion of the active stars without rotation
measurements does not likely effect our results, but may be explored
in a future study when we can ascertain the reason for the lack
of rotational modulation.

The fact that there are a small number of active, slowly rotating,
late-type M dwarfs in our sample suggests that the activity-rotation
relation may be more complicated than predicted by a simple spin-down
model.
%, particularly
% given the evidence that those same slowly rotating active stars appear
% to be drawn from a younger population. 
One possibility could be that
while persistent magnetic activity in late-type M dwarfs is rare (or
non-existent) for slow rotators, the presence of magnetic cycles may
produce observed activity in a small number of stars at any given
time.  Long-term observations are required to further test this
hypothesis. The small number of active, slowly rotating, late-type M
dwarfs could also be due to the young bias of MEarth, which might
exclude such stars.  An additional factor that could influence our analysis is
the potential presence of close binary companions, which have been
shown to affect the magnetic activity (and rotation) of stars
\citep[e.g.,][]{morgan12}.  While multiplicity likely plays a minor
role due to the small binary fraction of M dwarfs \citep{fischer92},
high resolution imaging and/or spectroscopy would be required to
further investigate its role.

Our analysis divided M dwarfs into early (M1-M4) and late-types
(M5-M8) and did not explore the ramifcations of altering the boundary
between the two populations.  Ideally, we would conduct the analysis
as a function of spectral type but we did not have enough stars in
each spectral type to produce a statistically significant result.  As
stated above, our analysis is therefore mostly a comparison between
M3-M4 and M5-M6 dwarfs.  Future and ongoing studies will
provide additional time-series photometric data with which we can
further study the rotation and activity behavior of M dwarfs (over a
larger range of spectral types), both surveys such as MEarth
\citep[both North and South;][]{irwin14}, APACHE \citep{sozzetti13},
SPECULOOS \citep{gillon13}, and
ExTrA\footnote{\href {http://www.eso.org/sci/meetings/2014/exoelt2014/presentations/Bonfils.pdf}{http://www.eso.org/sci/meetings/2014/exoelt2014/}\\
  \href{http://www.eso.org/sci/meetings/2014/exoelt2014/presentations/Bonfils.pdf}{presentations/Bonfils.pdf}} that focus specifically on M dwarfs and more general photometric surveys like Pan-STARRS \citep{pan-starrs} and LSST \citep{ivezic08}.

\acknowledgments The authors acknowledge Elisabeth Newton, Dylan
Morgan, and the other members of the Boston Area Drinking And Society
for Stars of Elfin Stature for useful conversations in the preparation
of this manuscript.  AAW acknowledges the support of NSF grants
AST-1109273 and AST-1255568 and the Research Corporation for Science
Advancement's Cottrell Scholarship.  KLW acknowledges the support of
the Boston University UROP program and the Clare Boothe Luce
scholarship. ZKBT gratefully acknowledges support from the Torres
Fellowship for Exoplanetary Research. The MEarth Team gratefully
acknowledges funding from the David and Lucille Packard Fellowship for
Science and Engineering (awarded to D.C.). This material is based upon
work supported by the National Science Foundation under grants
AST-0807690, AST-1109468, and AST-1004488 (Alan T. Waterman
Award). This publication was made possible through the support of a
grant from the John Templeton Foundation. The opinions expressed in
this publication are those of the authors and do not necessarily
reflect the views of the John Templeton Foundation.

\bibliographystyle{apj}

%\begin{thebibliography}

%\bibliography{ms}

\begin{thebibliography}{72}
%\expandafter\ifx\csname natexlab\endcsname\relax\def\natexlab#1{#1}\fi

\bibitem[{{Barnes}(2003)}]{barnes03}
{Barnes}, S.~A. 2003, \apj, 586, 464

\bibitem[{{Barry}(1988)}]{barry88}
{Barry}, D.~C. 1988, \apj, 334, 436

\bibitem[{{Bell} {et~al.}(2012){Bell}, {Hilton}, {Davenport}, {Hawley}, {West},
  \& {Rogel}}]{bell12}
{Bell}, K.~J., {Hilton}, E.~J., {Davenport}, J.~R.~A., {Hawley}, S.~L., {West},
  A.~A., \& {Rogel}, A.~B. 2012, \pasp, 124, 14

\bibitem[{{Berger} {et~al.}(2008){Berger}, {Gizis}, {Giampapa}, {Rutledge},
  {Liebert}, {Mart{\'{\i}}n}, {Basri}, {Fleming}, {Johns-Krull}, {Phan-Bao}, \&
  {Sherry}}]{berger08}
{Berger}, E., {et~al.} 2008, \apj, 673, 1080

\bibitem[{{Berta} {et~al.}(2012){Berta}, {Irwin}, {Charbonneau}, {Burke}, \&
  {Falco}}]{berta12}
{Berta}, Z.~K., {Irwin}, J., {Charbonneau}, D., {Burke}, C.~J., \& {Falco},
  E.~E. 2012, \aj, 144, 145

\bibitem[{{Bochanski} {et~al.}(2007{\natexlab{a}}){Bochanski}, {Munn},
  {Hawley}, {West}, {Covey}, \& {Schneider}}]{bochUVW}
{Bochanski}, J.~J., {Munn}, J.~A., {Hawley}, S.~L., {West}, A.~A., {Covey},
  K.~R., \& {Schneider}, D.~P. 2007{\natexlab{a}}, \aj, 134, 2418

\bibitem[{{Bochanski} {et~al.}(2007{\natexlab{b}}){Bochanski}, {West},
  {Hawley}, \& {Covey}}]{bochtem}
{Bochanski}, J.~J., {West}, A.~A., {Hawley}, S.~L., \& {Covey}, K.~R.
  2007{\natexlab{b}}, \aj, 133, 531

\bibitem[{{Browning}(2008)}]{browning08}
{Browning}, M.~K. 2008, \apj, 676, 1262

\bibitem[{{Browning} {et~al.}(2010){Browning}, {Basri}, {Marcy}, {West}, \&
  {Zhang}}]{browning10}
{Browning}, M.~K., {Basri}, G., {Marcy}, G.~W., {West}, A.~A., \& {Zhang}, J.
  2010, \aj, 139, 504

\bibitem[{{Burgasser} {et~al.}(2002){Burgasser}, {Liebert}, {Kirkpatrick}, \&
  {Gizis}}]{burgasser02}
{Burgasser}, A.~J., {Liebert}, J., {Kirkpatrick}, J.~D., \& {Gizis}, J.~E.
  2002, \aj, 123, 2744

\bibitem[{{Chabrier} \& {Baraffe}(1997)}]{chabrier97}
{Chabrier}, G., \& {Baraffe}, I. 1997, \aap, 327, 1039

\bibitem[{{Charbonneau} {et~al.}(2009)}]{Charbonneau09}
{Charbonneau}, D., {et~al.} 2009, \nat, 462, 891

\bibitem[{{Covey} {et~al.}(2007)}]{covey07}
{Covey}, K.~R., {et~al.} 2007, \aj, 134, 2398

\bibitem[{{Delfosse} {et~al.}(1998){Delfosse}, {Forveille}, {Perrier}, \&
  {Mayor}}]{delfosse98}
{Delfosse}, X., {Forveille}, T., {Perrier}, C., \& {Mayor}, M. 1998, \aap, 331,
  581

\bibitem[{{Dittmann} {et~al.}(2014){Dittmann}, {Irwin}, {Charbonneau}, \&
  {Berta-Thompson}}]{dittmann14}
{Dittmann}, J.~A., {Irwin}, J.~M., {Charbonneau}, D., \& {Berta-Thompson},
  Z.~K. 2014, \apj, 784, 156

\bibitem[{{Dobler} {et~al.}(2006){Dobler}, {Stix}, \& {Brandenburg}}]{dobler06}
{Dobler}, W., {Stix}, M., \& {Brandenburg}, A. 2006, \apj, 638, 336

\bibitem[{{Dressing} \& {Charbonneau}(2013)}]{dressing13}
{Dressing}, C.~D., \& {Charbonneau}, D. 2013, \apj, 767, 95

\bibitem[{{Dressing} \& {Charbonneau}(2015)}]{dressing15}
---. 2015, \apj, 807, 45

\bibitem[{{Eggen} \& {Iben}(1989)}]{eggen89}
{Eggen}, O.~J., \& {Iben}, Jr., I. 1989, \aj, 97, 431

\bibitem[{{Fischer} \& {Marcy}(1992)}]{fischer92}
{Fischer}, D.~A., \& {Marcy}, G.~W. 1992, \apj, 396, 178

\bibitem[{{Gillon} {et~al.}(2013){Gillon}, {Jehin}, {Delrez}, {Magain},
  {Opitom}, \& {Sohy}}]{gillon13}
{Gillon}, M., {Jehin}, E., {Delrez}, L., {Magain}, P., {Opitom}, C., \& {Sohy},
  S. 2013, in Protostars and Planets VI Posters, 66

\bibitem[{{Gizis} {et~al.}(2000){Gizis}, {Monet}, {Reid}, {Kirkpatrick},
  {Liebert}, \& {Williams}}]{gizis00}
{Gizis}, J.~E., {Monet}, D.~G., {Reid}, I.~N., {Kirkpatrick}, J.~D., {Liebert},
  J., \& {Williams}, R.~J. 2000, \aj, 120, 1085

\bibitem[{{Hawley} {et~al.}(1996){Hawley}, {Gizis}, \& {Reid}}]{hawley96}
{Hawley}, S.~L., {Gizis}, J.~E., \& {Reid}, I.~N. 1996, \aj, 112, 2799

\bibitem[{{Hawley} {et~al.}(1999){Hawley}, {Tourtellot}, \& {Reid}}]{hawley99}
{Hawley}, S.~L., {Tourtellot}, J.~G., \& {Reid}, I.~N. 1999, \aj, 117, 1341

\bibitem[{{Irwin} {et~al.}(2009{\natexlab{a}}){Irwin}, {Aigrain}, {Bouvier},
  {Hebb}, {Hodgkin}, {Irwin}, \& {Moraux}}]{irwin09}
{Irwin}, J., {Aigrain}, S., {Bouvier}, J., {Hebb}, L., {Hodgkin}, S., {Irwin},
  M., \& {Moraux}, E. 2009{\natexlab{a}}, \mnras, 392, 1456

\bibitem[{{Irwin} {et~al.}(2011{\natexlab{a}}){Irwin}, {Berta}, {Burke},
  {Charbonneau}, {Nutzman}, {West}, \& {Falco}}]{irwinrot}
{Irwin}, J., {Berta}, Z.~K., {Burke}, C.~J., {Charbonneau}, D., {Nutzman}, P.,
  {West}, A.~A., \& {Falco}, E.~E. 2011{\natexlab{a}}, \apj, 727, 56

\bibitem[{{Irwin} {et~al.}(2009{\natexlab{b}})}]{irwinMearth}
{Irwin}, J., {et~al.} 2009{\natexlab{b}}, \apj, 701, 1436

\bibitem[{{Irwin} {et~al.}(2014){Irwin}, {Berta-Thompson}, {Charbonneau},
  {Dittmann}, {Falco}, {Newton}, \& {Nutzman}}]{irwin14}
{Irwin}, J.~M., {Berta-Thompson}, Z.~K., {Charbonneau}, D., {Dittmann}, J.,
  {Falco}, E.~E., {Newton}, E.~R., \& {Nutzman}, P. 2014, ArXiv e-prints

\bibitem[{{Irwin} {et~al.}(2011{\natexlab{b}})}]{irwin11}
{Irwin}, J.~M., {et~al.} 2011{\natexlab{b}}, \apj, 742, 123

\bibitem[{{Ivezic} {et~al.}(2008){Ivezic}, {Tyson}, {Abel}, {Acosta},
  {Allsman}, {AlSayyad}, {Anderson}, {Andrew}, {Angel}, {Angeli}, {Ansari},
  {Antilogus}, {Arndt}, {Astier}, {Aubourg}, {Axelrod}, {Bard}, {Barr},
  {Barrau}, {Bartlett}, {Bauman}, {Beaumont}, {Becker}, {Becla}, {Beldica},
  {Bellavia}, {Blanc}, {Blandford}, {Bloom}, {Bogart}, {Borne}, {Bosch},
  {Boutigny}, {Brandt}, {Brown}, {Bullock}, {Burchat}, {Burke}, {Cagnoli},
  {Calabrese}, {Chandrasekharan}, {Chesley}, {Cheu}, {Chiang}, {Claver},
  {Connolly}, {Cook}, {Cooray}, {Covey}, {Cribbs}, {Cui}, {Cutri}, {Daubard},
  {Daues}, {Delgado}, {Digel}, {Doherty}, {Dubois}, {Dubois-Felsmann},
  {Durech}, {Eracleous}, {Ferguson}, {Frank}, {Freemon}, {Gangler}, {Gawiser},
  {Geary}, {Gee}, {Geha}, {Gibson}, {Gilmore}, {Glanzman}, {Goodenow},
  {Gressler}, {Gris}, {Guyonnet}, {Hascall}, {Haupt}, {Hernandez}, {Hogan},
  {Huang}, {Huffer}, {Innes}, {Jacoby}, {Jain}, {Jee}, {Jernigan},
  {Jevremovic}, {Johns}, {Jones}, {Juramy-Gilles}, {Juric}, {Kahn}, {Kalirai},
  {Kallivayalil}, {Kalmbach}, {Kantor}, {Kasliwal}, {Kessler}, {Kirkby},
  {Knox}, {Kotov}, {Krabbendam}, {Krughoff}, {Kubanek}, {Kuczewski},
  {Kulkarni}, {Lambert}, {Le Guillou}, {Levine}, {Liang}, {Lim}, {Lintott},
  {Lupton}, {Mahabal}, {Marshall}, {Marshall}, {May}, {McKercher}, {Migliore},
  {Miller}, {Mills}, {Monet}, {Moniez}, {Neill}, {Nief}, {Nomerotski},
  {Nordby}, {O'Connor}, {Oliver}, {Olivier}, {Olsen}, {Ortiz}, {Owen}, {Pain},
  {Peterson}, {Petry}, {Pierfederici}, {Pietrowicz}, {Pike}, {Pinto}, {Plante},
  {Plate}, {Price}, {Prouza}, {Radeka}, {Rajagopal}, {Rasmussen}, {Regnault},
  {Ridgway}, {Ritz}, {Rosing}, {Roucelle}, {Rumore}, {Russo}, {Saha},
  {Sassolas}, {Schalk}, {Schindler}, {Schneider}, {Schumacher}, {Sebag},
  {Sembroski}, {Seppala}, {Shipsey}, {Silvestri}, {Smith}, {Smith}, {Strauss},
  {Stubbs}, {Sweeney}, {Szalay}, {Takacs}, {Thaler}, {Van Berg}, {Vanden Berk},
  {Vetter}, {Virieux}, {Xin}, {Walkowicz}, {Walter}, {Wang}, {Warner},
  {Willman}, {Wittman}, {Wolff}, {Wood-Vasey}, {Yoachim}, {Zhan}, \& {for the
  LSST Collaboration}}]{ivezic08}
{Ivezic}, Z., {et~al.} 2008, ArXiv e-prints

\bibitem[{{Kaiser} {et~al.}(2002)}]{pan-starrs}
{Kaiser}, N., {et~al.} 2002, in Society of Photo-Optical Instrumentation
  Engineers (SPIE) Conference Series, Vol. 4836, Survey and Other Telescope
  Technologies and Discoveries, ed. J.~A. {Tyson} \& S.~{Wolff}, 154--164

\bibitem[{{Kiraga} \& {Stepien}(2007)}]{kiraga07}
{Kiraga}, M., \& {Stepien}, K. 2007, \actaa, 57, 149

\bibitem[{Koenker(2015)}]{quantreg}
Koenker, R. 2015, quantreg: Quantile Regression, r package version 5.11

\bibitem[{{Law} {et~al.}(2012)}]{law12}
{Law}, N.~M., {et~al.} 2012, \apj, 757, 133

\bibitem[{{Lee} {et~al.}(2010){Lee}, {Berger}, \& {Knapp}}]{lee12}
{Lee}, K.-G., {Berger}, E., \& {Knapp}, G.~R. 2010, \apj, 708, 1482

\bibitem[{{Leggett}(1992)}]{leggett92}
{Leggett}, S.~K. 1992, \apjs, 82, 351

\bibitem[{{L{\'e}pine} {et~al.}(2005){L{\'e}pine}, {Rich}, \&
  {Shara}}]{lepine05}
{L{\'e}pine}, S., {Rich}, R.~M., \& {Shara}, M.~M. 2005, \apjl, 633, L121

\bibitem[{{L{\'e}pine} \& {Shara}(2005)}]{lspm}
{L{\'e}pine}, S., \& {Shara}, M.~M. 2005, \aj, 129, 1483

\bibitem[{{Mamajek} \& {Hillenbrand}(2008)}]{mamajek08}
{Mamajek}, E.~E., \& {Hillenbrand}, L.~A. 2008, \apj, 687, 1264

\bibitem[{{Mohanty} \& {Basri}(2003)}]{mohanty03}
{Mohanty}, S., \& {Basri}, G. 2003, \apj, 583, 451

\bibitem[{{Montes} {et~al.}(2001){Montes}, {L{\'o}pez-Santiago}, {G{\'a}lvez},
  {Fern{\'a}ndez-Figueroa}, {De Castro}, \& {Cornide}}]{montes01}
{Montes}, D., {L{\'o}pez-Santiago}, J., {G{\'a}lvez}, M.~C.,
  {Fern{\'a}ndez-Figueroa}, M.~J., {De Castro}, E., \& {Cornide}, M. 2001,
  \mnras, 328, 45

\bibitem[{{Morgan} {et~al.}(2012){Morgan}, {West}, {Garc{\'e}s}, {Catal{\'a}n},
  {Dhital}, {Fuchs}, \& {Silvestri}}]{morgan12}
{Morgan}, D.~P., {West}, A.~A., {Garc{\'e}s}, A., {Catal{\'a}n}, S., {Dhital},
  S., {Fuchs}, M., \& {Silvestri}, N.~M. 2012, \aj, 144, 93

\bibitem[{{Muirhead} {et~al.}(2012){Muirhead}, {Hamren}, {Schlawin},
  {Rojas-Ayala}, {Covey}, \& {Lloyd}}]{muirhead12}
{Muirhead}, P.~S., {Hamren}, K., {Schlawin}, E., {Rojas-Ayala}, B., {Covey},
  K.~R., \& {Lloyd}, J.~P. 2012, \apjl, 750, L37

\bibitem[{{Newton} {et~al.}(2014){Newton}, {Charbonneau}, {Irwin},
  {Berta-Thompson}, {Rojas-Ayala}, {Covey}, \& {Lloyd}}]{newton14}
{Newton}, E.~R., {Charbonneau}, D., {Irwin}, J., {Berta-Thompson}, Z.~K.,
  {Rojas-Ayala}, B., {Covey}, K., \& {Lloyd}, J.~P. 2014, \aj, 147, 20

\bibitem[{{Nidever} {et~al.}(2002){Nidever}, {Marcy}, {Butler}, {Fischer}, \&
  {Vogt}}]{nidev02}
{Nidever}, D.~L., {Marcy}, G.~W., {Butler}, R.~P., {Fischer}, D.~A., \& {Vogt},
  S.~S. 2002, \apjs, 141, 503

\bibitem[{{Nutzman} \& {Charbonneau}(2008)}]{nutzman08}
{Nutzman}, P., \& {Charbonneau}, D. 2008, \pasp, 120, 317

\bibitem[{{Pizzolato} {et~al.}(2003){Pizzolato}, {Maggio}, {Micela},
  {Sciortino}, \& {Ventura}}]{pizzolato03}
{Pizzolato}, N., {Maggio}, A., {Micela}, G., {Sciortino}, S., \& {Ventura}, P.
  2003, \aap, 397, 147

\bibitem[{{R Core Team}(2014)}]{r}
{R Core Team}. 2014, R: A Language and Environment for Statistical Computing, R
  Foundation for Statistical Computing, Vienna, Austria

\bibitem[{{Reid} \& {Hawley}(2005)}]{reid05}
{Reid}, I.~N., \& {Hawley}, S.~L. 2005, {New light on dark stars : red dwarfs,
  low-mass stars, brown dwarfs} (Springer-Praxis books in astrophysics and
  astronomy.~Praxis Publishing Ltd)

\bibitem[{{Reid} {et~al.}(1995){Reid}, {Hawley}, \& {Gizis}}]{reid95}
{Reid}, I.~N., {Hawley}, S.~L., \& {Gizis}, J.~E. 1995, \aj, 110, 1838

\bibitem[{{Reiners} \& {Basri}(2008)}]{reiners08}
{Reiners}, A., \& {Basri}, G. 2008, \apj, 684, 1390

\bibitem[{{Reiners} \& {Basri}(2009)}]{reiners09}
---. 2009, \apj, 705, 1416

\bibitem[{{Reiners} \& {Mohanty}(2012)}]{reiners12}
{Reiners}, A., \& {Mohanty}, S. 2012, \apj, 746, 43

\bibitem[{{Sch{\"o}nrich} {et~al.}(2010){Sch{\"o}nrich}, {Binney}, \&
  {Dehnen}}]{schonrich10}
{Sch{\"o}nrich}, R., {Binney}, J., \& {Dehnen}, W. 2010, \mnras, 403, 1829

\bibitem[{{Shkolnik} {et~al.}(2010){Shkolnik}, {Hebb}, {Liu}, {Reid}, \&
  {Collier Cameron}}]{shkolnik10}
{Shkolnik}, E.~L., {Hebb}, L., {Liu}, M.~C., {Reid}, I.~N., \& {Collier
  Cameron}, A. 2010, \apj, 716, 1522

\bibitem[{{Silvestri} {et~al.}(2005){Silvestri}, {Hawley}, \&
  {Oswalt}}]{silvestri05}
{Silvestri}, N.~M., {Hawley}, S.~L., \& {Oswalt}, T.~D. 2005, \aj, 129, 2428

\bibitem[{{Skumanich}(1972)}]{skumanich72}
{Skumanich}, A. 1972, \apj, 171, 565

\bibitem[{{Soderblom}(2010)}]{soderblom10}
{Soderblom}, D.~R. 2010, \araa, 48, 581

\bibitem[{{Soderblom} {et~al.}(1991){Soderblom}, {Duncan}, \&
  {Johnson}}]{soderblom91}
{Soderblom}, D.~R., {Duncan}, D.~K., \& {Johnson}, D.~R.~H. 1991, \apj, 375,
  722

\bibitem[{{Sozzetti} {et~al.}(2013){Sozzetti}, {Bernagozzi}, {Bertolini},
  {Calcidese}, {Carbognani}, {Cenadelli}, {Christille}, {Damasso}, {Giacobbe},
  {Lanteri}, {Lattanzi}, \& {Smart}}]{sozzetti13}
{Sozzetti}, A., {et~al.} 2013, in European Physical Journal Web of Conferences,
  Vol.~47, European Physical Journal Web of Conferences, 3006

\bibitem[{{Stauffer} {et~al.}(1994){Stauffer}, {Caillault}, {Gagne}, {Prosser},
  \& {Hartmann}}]{stauffer94}
{Stauffer}, J.~R., {Caillault}, J.-P., {Gagne}, M., {Prosser}, C.~F., \&
  {Hartmann}, L.~W. 1994, \apjs, 91, 625

\bibitem[{{Stauffer} \& {Hartmann}(1986)}]{stauffer86}
{Stauffer}, J.~R., \& {Hartmann}, L.~W. 1986, \apjs, 61, 531

\bibitem[{{Walkowicz} \& {Hawley}(2009)}]{walkowicz09}
{Walkowicz}, L.~M., \& {Hawley}, S.~L. 2009, \aj, 137, 3297

\bibitem[{{Walkowicz} {et~al.}(2004){Walkowicz}, {Hawley}, \&
  {West}}]{walkowicz04}
{Walkowicz}, L.~M., {Hawley}, S.~L., \& {West}, A.~A. 2004, \pasp, 116, 1105

\bibitem[{{West} \& {Basri}(2009)}]{westbasri09}
{West}, A.~A., \& {Basri}, G. 2009, \apj, 693, 1283

\bibitem[{{West} {et~al.}(2006){West}, {Bochanski}, {Hawley}, {Cruz}, {Covey},
  {Silvestri}, {Reid}, \& {Liebert}}]{west06}
{West}, A.~A., {Bochanski}, J.~J., {Hawley}, S.~L., {Cruz}, K.~L., {Covey},
  K.~R., {Silvestri}, N.~M., {Reid}, I.~N., \& {Liebert}, J. 2006, \aj, 132,
  2507

\bibitem[{{West} {et~al.}(2008){West}, {Hawley}, {Bochanski}, {Covey}, {Reid},
  {Dhital}, {Hilton}, \& {Masuda}}]{west08}
{West}, A.~A., {Hawley}, S.~L., {Bochanski}, J.~J., {Covey}, K.~R., {Reid},
  I.~N., {Dhital}, S., {Hilton}, E.~J., \& {Masuda}, M. 2008, \aj, 135, 785

\bibitem[{{West} {et~al.}(2004)}]{west04}
{West}, A.~A., {et~al.} 2004, \aj, 128, 426

\bibitem[{{West} {et~al.}(2011)}]{west11}
---. 2011, \aj, 141, 97

\bibitem[{{Williams} {et~al.}(2014){Williams}, {Berger}, {Irwin},
  {Berta-Thompson}, \& {Charbonneau}}]{williams.2014.smomaudiaybn32j}
{Williams}, P.~K.~G., {Berger}, E., {Irwin}, J., {Berta-Thompson}, Z.~K., \&
  {Charbonneau}, D. 2014, ArXiv e-prints

\bibitem[{{Wilson} \& {Woolley}(1970)}]{wilson70}
{Wilson}, O., \& {Woolley}, R. 1970, \mnras, 148, 463

\bibitem[{{York} {et~al.}(2000)}]{york00}
{York}, D.~G., {et~al.} 2000, \aj, 120, 1579

\end{thebibliography}
%\end{thebibliography}

\clearpage
\begin{center}
\begin{turnpage}
\begin{deluxetable*}{lcccccrrrrrrrrcc}
%\rotate
%\landscape
\tablewidth{0pt}
\tablewidth{0pt}
\tablecolumns{15}
\tabletypesize{\tiny}
\tablecaption{Kinematics, Activity, and Rotation Periods for MEarth M Dwarfs\tablenotemark{a}}
\renewcommand{\arraystretch}{0}
\tablehead{
\colhead{Name}&
\colhead{RA}&
\colhead{Dec}&
\colhead{Sp.}&
\colhead{Dist\tablenotemark{a}}&
\colhead{Dist\tablenotemark{a}}&
\colhead{RV}&
\colhead{PM$_{\rm RA}$\tablenotemark{b}}&
\colhead{PM$_{\rm Dec}$\tablenotemark{b}}&
\colhead{$U$\tablenotemark{c}}&
\colhead{$V$\tablenotemark{c}}&
\colhead{$W$\tablenotemark{c}}&
\colhead{Per.}&
\colhead{H$\alpha$ EW\tablenotemark{d}}&
\colhead{Act.}&
\colhead{Rot.}\\
\colhead{}&
\colhead{}&
\colhead{}&
\colhead{Type}&
\colhead{(pc)}&
\colhead{Flag}&
\colhead{(km/s)}&
\colhead{(mas/yr)}&
\colhead{(mas/yr)}&
\colhead{(km/s)}&
\colhead{(km/s)}&
\colhead{(km/s)}&
\colhead{(days)}&
\colhead{(\AA)}&
\colhead{Flag\tablenotemark{e}}&
\colhead{Flag\tablenotemark{f}}
}
\startdata
LSPMJ0001+0659    & 00:01:15.809 & +06:59:35.65 & M6 & $   16.7$ & D & $   -0.7$ & $ -447$ & $  -81$ & $  +11.1 \pm 1.5    $ & $   +0.1 \pm 8.4    $ & $  +13.8 \pm 4.6    $ & $   20.4$ & $   +6.9 \pm 1.2    $ & $1$ & $1$\\ 
LSPMJ0015+4344    & 00:15:18.827 & +43:44:34.72 & M5 & $   25.7$ & L & $   +5.9$ & $ +232$ & $  +38$ & $  +10.1 \pm 4.1    $ & $   +4.6 \pm 8.3    $ & $   +7.1 \pm 2.8    $ & $    1.4$ & $   +6.5 \pm 0.8    $ & $1$ & $1$\\ 
LSPMJ0015+1333    & 00:15:49.244 & +13:33:22.25 & M3 & $    9.6$ & D & $  +65.1$ & $ +621$ & $ +333$ & $   +0.2 \pm 2.2    $ & $  +32.1 \pm 8.3    $ & $  -25.4 \pm 4.3    $ &  \nodata  &        \nodata        & $0$ & $0$\\ 
LSPMJ0016+1951E   & 00:16:16.141 & +19:51:50.56 & M4 & $   22.2$ & L & $   +9.0$ & $ +709$ & $ -748$ & $  +10.3 \pm 2.8    $ & $   +4.1 \pm 8.3    $ & $   +7.9 \pm 4.1    $ &  \nodata  &        \nodata        & $0$ & $0$\\ 
LSPMJ0016+2003    & 00:16:56.803 & +20:03:55.17 & M4 & $   22.2$ & L & $  +12.0$ & $ +228$ & $  +24$ & $   +9.2 \pm 2.8    $ & $   +6.9 \pm 8.3    $ & $   +5.2 \pm 4.0    $ & $   17.3$ & $   +3.5 \pm 0.4    $ & $1$ & $1$\\ 
LSPMJ0018+2748    & 00:18:53.590 & +27:48:49.81 & M4 & $   19.2$ & L & $  +12.7$ & $ +387$ & $ -101$ & $   +8.5 \pm 3.3    $ & $   +8.3 \pm 8.3    $ & $   +4.5 \pm 3.6    $ & $    6.0$ & $   +1.6 \pm 0.3    $ & $1$ & $1$\\ 
LSPMJ0024+2626    & 00:24:03.799 & +26:26:29.76 & M4 & $   26.1$ & D & $  +16.0$ & $ +162$ & $  -55$ & $   +8.3 \pm 3.5    $ & $   +8.5 \pm 8.3    $ & $   +4.2 \pm 3.7    $ & $   29.9$ & $   +1.3 \pm 0.4    $ & $1$ & $1$\\ 
LSPMJ0024+3002    & 00:24:34.876 & +30:02:29.59 & M5 & $   15.5$ & D & $  +17.6$ & $ +580$ & $  +28$ & $   +7.1 \pm 3.6    $ & $  +10.8 \pm 8.3    $ & $   +2.9 \pm 3.4    $ & $    1.1$ & $   +6.6 \pm 0.9    $ & $1$ & $1$\\ 
LSPMJ0028+5022    & 00:28:53.972 & +50:22:33.17 & M4 & $   16.9$ & D & $  +24.2$ & $ +423$ & $ +124$ & $   +2.8 \pm 4.6    $ & $  +17.8 \pm 8.3    $ & $   +3.8 \pm 1.8    $ & $    1.1$ & $   +5.9 \pm 0.6    $ & $1$ & $1$\\ 
LSPMJ0033+1448    & 00:33:22.350 & +14:48:06.45 & M5 & $   27.3$ & D & $  +21.5$ & $ +278$ & $  -71$ & $   +6.7 \pm 3.1    $ & $  +11.6 \pm 8.2    $ & $   -1.2 \pm 4.2    $ & $    0.4$ & $   +3.5 \pm 0.6    $ & $1$ & $2$\\ 
LSPMJ0035+5241N   & 00:35:53.680 & +52:41:36.59 & M4 &  \nodata  & L & $  +12.7$ & $ +787$ & $ -186$ &        \nodata        &        \nodata        &        \nodata        &  \nodata  &        \nodata        & $0$ & $0$\\ 
LSPMJ0038+6150    & 00:38:27.677 & +61:50:06.34 & M4 & $   46.5$ & D & $  -58.3$ & $ +349$ & $  -43$ & $  +37.5 \pm 5.1    $ & $  -40.1 \pm 8.2    $ & $   +7.9 \pm 1.2    $ & $   62.4$ &        \nodata        & $0$ & $2$\\ 
LSPMJ0039+1454S   & 00:39:33.544 & +14:54:18.96 & M4 &  \nodata  & L & $   +9.1$ & $ +321$ & $  +39$ &        \nodata        &        \nodata        &        \nodata        & $   34.0$ &        \nodata        & $0$ & $1$\\ 
LSPMJ0103+6221    & 01:03:19.824 & +62:21:55.74 & M6 & $   10.5$ & L & $   +6.7$ & $ +739$ & $  +86$ & $   +9.3 \pm 5.4    $ & $   +5.9 \pm 7.9    $ & $   +7.0 \pm 0.5    $ & $    1.0$ & $   +9.2 \pm 1.3    $ & $1$ & $1$\\ 
LSPMJ0130+0236    & 01:30:43.122 & +02:36:37.01 & M6 & $   18.3$ & D & $  -15.6$ & $ -119$ & $ -194$ & $  +18.3 \pm 4.1    $ & $   -1.5 \pm 7.6    $ & $  +24.3 \pm 4.6    $ &  \nodata  & $   +0.8 \pm 0.5    $ & $1$ & $0$\\ 
LSPMJ0153+0147    & 01:53:30.754 & +01:47:55.89 & M6 & $   20.1$ & D & $  +30.4$ & $ +427$ & $  +45$ & $   +1.4 \pm 4.7    $ & $   +9.4 \pm 7.2    $ & $   -8.2 \pm 4.6    $ & $    0.2$ & $   +8.6 \pm 1.2    $ & $1$ & $1$\\ 
LSPMJ0153+4427    & 01:53:49.551 & +44:27:28.49 & M5 & $   20.1$ & D & $  +14.3$ & $ +245$ & $  -92$ & $   +6.4 \pm 6.5    $ & $   +8.6 \pm 7.1    $ & $   +5.4 \pm 0.7    $ & $    0.2$ & $  +17.8 \pm 2.3    $ & $1$ & $1$\\ 
LSPMJ0158+4049    & 01:58:45.211 & +40:49:44.51 & M5 & $   17.1$ & D & $  +19.1$ & $ +399$ & $  -69$ & $   +3.5 \pm 6.5    $ & $  +11.1 \pm 7.0    $ & $   +3.7 \pm 0.9    $ & $    0.5$ & $   +7.3 \pm 1.0    $ & $1$ & $1$\\ 
LSPMJ0202+1334    & 02:02:44.355 & +13:34:33.21 & M5 & $   20.7$ & D & $  +15.6$ & $ +461$ & $ -111$ & $   +6.7 \pm 5.7    $ & $   +7.0 \pm 7.0    $ & $   +3.0 \pm 3.5    $ & $    4.0$ & $   +7.6 \pm 0.9    $ & $1$ & $1$\\ 
LSPMJ0212+0000    & 02:12:54.624 & +00:00:16.71 & M4 & $    9.8$ & D & $  +33.0$ & $ +560$ & $  +34$ & $   +0.5 \pm 5.1    $ & $   +8.0 \pm 6.7    $ & $   -8.1 \pm 4.7    $ & $    4.7$ & $   +2.2 \pm 0.4    $ & $1$ & $1$\\ 

\nodata & \nodata   & \nodata   & \nodata   & \nodata   & \nodata   & \nodata   & \nodata   & \nodata   & \nodata   & \nodata   & \nodata   & \nodata  & \nodata   & \nodata  & \nodata  \\

\enddata

\tablenotetext{a}{A Distance Flag of "D" indicates the distance is a trigonometric parallax measurement derived from MEarth imaging, as published in \citet{dittmann14}. A flag of "L" indicates the distance comes from \citet{lepine05} and may be either photometric, spectroscopic, or trigonometric.}

\tablenotetext{b}{Proper motions are quoted as projected on the plane of the sky, with (PM$_{\rm RA}$, PM$_{\rm Dec}$) = $(\mu_{\alpha}\cos \delta, \mu_{\delta})$.}

\tablenotetext{c}{$UVW$ velocities are quoted in a right-handed coordinate system, with $U$ pointed towards the Galactic center.}

\tablenotetext{d}{This table uses a convention in which emission lines correspond to equivalent widths $>0$.}

\tablenotetext{e}{Active M dwarfs have an Active Flag value of 1; inactive stars have a flag of 0 (see Section \ref{sec:act}).}

\tablenotetext{f}{Stars with measured rotation periods have a Rotation Flag value of 1 or 2, with a value of 1 being more robust. Stars whose rotation periods did not cross our detection threshold are flagged as 0 (see Section \ref{sec:rot}). }

\tablecomments{This table is published in its entirety in the
  electronic edition of the Astrophysical Journal. A portion is shown
  here for guidance concerning its form and content. The
  machine-readable table also contains the semiamplitudes of the
  rotation data as well as other extracted parameters from the spectra used in this paper, including a calculation of $L_{\rm{H\alpha}}$/$L_{\rm{bol}}$ and atomic/molecular indices for Na, TiO, CaH, CaOH features \citep[as defined in][]{reid95}.}

\label{table:data}
\end{deluxetable*}
\end{turnpage}
\end{center}

\begin{center}
\begin{deluxetable*}{ccccccc}
\tablewidth{0pt}
\tablewidth{0pt}
\tablecolumns{7} 
%\tabletypesize{\scriptsize}
\tablecaption{Population Kinematics - Entire Sample}
\renewcommand{\arraystretch}{1}
\tablehead{
\colhead{Mean Rotation}&
\colhead{$\mu_U$}&
\colhead{$\mu_V$}&
\colhead{$\mu_W$}&
\colhead{$\sigma_U$}&
\colhead{$\sigma_V$}&
\colhead{$\sigma_W$}\\
\colhead{Period (days)}&
\colhead{(km\ s$^{-1}$)}&
\colhead{(km\ s$^{-1}$)}&
\colhead{(km\ s$^{-1}$)}&
\colhead{(km\ s$^{-1}$)}&
\colhead{(km\ s$^{-1}$)}&
\colhead{(km\ s$^{-1}$)}
}
\startdata
0.5&  9.8$_{1.3}^{0.3}$ & 3.9$_{0.8}^{0.7}$ & 5.0$_{0.9}^{1.0}$ & \nodata & 4.5$_{0.5}^{0.6}$ & 7.2$_{0.7}^{0.8}$\\
3.3 & 5.6$_{1.3}^{1.3}$ & 3.9$_{1.3}^{1.3}$ &  9.2$_{1.4}^{1.4}$ &
8.8$_{1.5}^{1.6}$ & 8.4$_{0.9}^{1.0}$ & 9.8$_{0.9}^{1.1}$\\
49.0 & 2.2$_{2.1}^{2.2}$ & -0.3$_{1.6}^{1.7}$ & 14.4$_{1.9}^{2.0}$ &
14.2$_{1.7}^{1.9}$ & 10.2$_{1.2}^{1.4}$ & 13.2$_{1.2}^{1.5}$\\
119.8 & 8.9$_{2.0}^{1.8}$ & -1.4$_{2.4}^{2.6}$ & 8.1$_{3.8}^{3.7}$ &
6.3$_{1.7}^{2.1}$ & 7.4$_{1.7}^{2.2}$ & 14.6$_{2.4}^{3.4}$\\
\enddata
\label{table:kinematics}
\end{deluxetable*}
\end{center}

% \begin{deluxetable}{ccccccc}
% \tablewidth{0pt}
% \tablewidth{0pt}
% \tablecolumns{7} 
% %\tabletypesize{\scriptsize}
% \tablecaption{Population Kinematics - Active M Dwarfs}
% \renewcommand{\arraystretch}{1}
% \tablehead{
% \colhead{Mean Rotation}&
% \colhead{$\mu_U$}&
% \colhead{$\mu_V$}&
% \colhead{$\mu_W$}&
% \colhead{$\sigma_U$}&
% \colhead{$\sigma_V$}&
% \colhead{$\sigma_W$}\\
% \colhead{Period (days)}&
% \colhead{(km\ s$^{-1}$)}&
% \colhead{(km\ s$^{-1}$)}&
% \colhead{(km\ s$^{-1}$)}&
% \colhead{(km\ s$^{-1}$)}&
% \colhead{(km\ s$^{-1}$)}&
% \colhead{(km\ s$^{-1}$)}}
% \startdata
% 0.5 &  9.7$_{0.3}^{0.3}$ & 3.7$_{0.8}^{0.7}$ & 5.6$_{0.9}^{1.0}$ & \nodata & 4.5$_{0.6}^{0.6}$ & 7.2$_{0.7}^{0.8}$\\
% 3.2 & 6.6$_{1.6}^{1.6}$ & 4.4$_{1.5}^{1.5}$ &  7.2$_{1.2}^{1.3}$ &
% 9.3$_{1.8}^{1.9}$ & 8.4$_{0.9}^{1.2}$ & 7.3$_{0.9}^{1.0}$\\
% 27.1 & 1.7$_{3.0}^{2.8}$ & 3.2$_{5.6}^{1.7}$ & 19.0$_{3.9}^{3.8}$ &
% 9.8$_{1.9}^{2.5}$ &\nodata & 14.1$_{2.3}^{3.1}$\\
% 119.5 & 11.9$_{1.7}^{1.7}$ & 4.6$_{4.0}^{4.1}$ & -5.3$_{8.4}^{8.6}$ &
% \nodata & \nodata & 11.3$_{4.0}^{7.2}$\\
% \enddata
% \label{table:kinematics_act}
% \end{deluxetable}

\end{document}